\documentclass [12pt] {article}
\usepackage{a4}

  \def\pp{{\mathchoice
          {
              \kern 1pt%
              \raise 1pt
              \vbox{\hrule width5pt height0.4pt depth0pt
                    \kern -2pt
                    \hbox{\kern 2.3pt
                          \vrule width0.4pt height6pt depth0pt
                          }
                    \kern -2pt
                    \hrule width5pt height0.4pt depth0pt}%
                    \kern 1pt
           }
            {
              \kern 1pt%
              \raise 1pt
              \vbox{\hrule width4.3pt height0.4pt depth0pt
                    \kern -1.8pt
                    \hbox{\kern 1.95pt
                          \vrule width0.4pt height5.4pt depth0pt
                          }
                    \kern -1.8pt
                    \hrule width4.3pt height0.4pt depth0pt}%
                    \kern 1pt
            }
            {
              \kern 0.5pt%
              \raise 1pt
              \vbox{\hrule width4.0pt height0.3pt depth0pt
                    \kern -1.9pt  
                    \hbox{\kern 1.85pt
                          \vrule width0.3pt height5.7pt depth0pt
                          }
                    \kern -1.9pt
                    \hrule width4.0pt height0.3pt depth0pt}%
                    \kern 0.5pt
            }
            {
              \kern 0.5pt%
              \raise 1pt
              \vbox{\hrule width3.6pt height0.3pt depth0pt
                    \kern -1.5pt
                    \hbox{\kern 1.65pt
                          \vrule width0.3pt height4.5pt depth0pt
                          }
                    \kern -1.5pt
                    \hrule width3.6pt height0.3pt depth0pt}%
                    \kern 0.5pt
            }
        }}

  \def\mm{{\mathchoice
                       {
                             \kern 1pt
               \raise 1pt    \vbox{\hrule width5pt height0.4pt depth0pt
                                  \kern 2pt
                                  \hrule width5pt height0.4pt depth0pt}
                             \kern 1pt}
                       {
                            \kern 1pt
               \raise 1pt \vbox{\hrule width4.3pt height0.4pt depth0pt
                                  \kern 1.8pt
                                  \hrule width4.3pt height0.4pt depth0pt}
                             \kern 1pt}
                       {
                            \kern 0.5pt
               \raise 1pt
                            \vbox{\hrule width4.0pt height0.3pt depth0pt
                                  \kern 1.9pt
                                  \hrule width4.0pt height0.3pt depth0pt}
                            \kern 1pt}
                       {
                           \kern 0.5pt
             \raise 1pt  \vbox{\hrule width3.6pt height0.3pt depth0pt
                                  \kern 1.5pt
                                  \hrule width3.6pt height0.3pt depth0pt}
                           \kern 0.5pt}
                       }}

\catcode`@=11
\def\un#1{\relax\ifmmode\@@underline#1\else
        $\@@underline{\hbox{#1}}$\relax\fi}
\catcode`@=12

\let\du=\du                     

\def\a{\alpha}
\def\b{\beta}

\def\d{\delta}

\def\f{\phi}
\def\g{\gamma}
\def\h{\eta}

\def\j{\psi}
\def\k{\kappa}

\def\m{\mu}
\def\n{\nu}

\def\p{\pi}
\def\q{\theta}
\def\r{\rho}

\def\t{\tau}

\def\z{\zeta}

\def\G{\Gamma}

\def\L{\Lambda}

\def\Q{\Theta}

\def\vf{\varphi}

\def\ch{{\cal H}}

\def\cm{{\cal M}}

\def\bo{{\raise-.5ex\hbox{\large$\Box$}}}               
\def\pa{\partial}                                       
\def\de{\nabla}                                         
\def\TH{{\raise.2ex\hbox{$\displaystyle \bigodot$}\mskip-4.7mu \llap H \;}}
\def\face{{\raise.2ex\hbox{$\displaystyle \bigodot$}\mskip-2.2mu \llap {$\ddot
        \smile$}}}                                      

\def\VEV#1{\left\langle #1\right\rangle}        
\def\abs#1{\left| #1\right|}                    
\def\leftrightarrowfill{$\mathsurround=0pt \mathord\leftarrow \mkern-6mu
        \cleaders\hbox{$\mkern-2mu \mathord- \mkern-2mu$}\hfill
        \mkern-6mu \mathord\rightarrow$}
\def\dvec#1{\vbox{\ialign{##\crcr
        \leftrightarrowfill\crcr\noalign{\kern-1pt\nointerlineskip}
        $\hfil\displaystyle{#1}\hfil$\crcr}}}           

\def\frac#1#2{{\textstyle{#1\over\vphantom2\smash{\raise.20ex
        \hbox{$\scriptstyle{#2}$}}}}}                   
\def\sfrac#1#2{{\vphantom1\smash{\lower.5ex\hbox{\small$#1$}}\over
        \vphantom1\smash{\raise.4ex\hbox{\small$#2$}}}} 
\def\bfrac#1#2{{\vphantom1\smash{\lower.5ex\hbox{$#1$}}\over
        \vphantom1\smash{\raise.3ex\hbox{$#2$}}}}       
\def\afrac#1#2{{\vphantom1\smash{\lower.5ex\hbox{$#1$}}\over#2}}    

\def\[{\lfloor{\hskip 0.35pt}\!\!\!\lceil}
\def\]{\rfloor{\hskip 0.35pt}\!\!\!\rceil}
\def\Lag{{\cal L}}
\def\du#1#2{_{#1}{}^{#2}}

\def\fracm#1#2{\hbox{\large{${\frac{{#1}}{{#2}}}$}}}

\def\un{\underline}
\def\fracmm#1#2{{{#1}\over{#2}}}

\def\low#1{{\raise -3pt\hbox{${\hskip 0.75pt}\!_{#1}$}}}

\newskip\humongous \humongous=0pt plus 1000pt minus 1000pt
\def\caja{\mathsurround=0pt}
\def\eqalign#1{\,\vcenter{\openup2\jot \caja
        \ialign{\strut \hfil$\displaystyle{##}$&$
        \displaystyle{{}##}$\hfil\crcr#1\crcr}}\,}
\newif\ifdtup

\def\pl#1#2#3{Phys.~Lett.~{\bf {#1}B} (19{#2}) #3}
\def\np#1#2#3{Nucl.~Phys.~{\bf B{#1}} (19{#2}) #3}

\def\cqg#1#2#3{Class.~and Quantum Grav.~{\bf {#1}} (19{#2}) #3}

\parskip=\medskipamount                 
\thicklines                         

\begin{document}
\thispagestyle{empty}

{\hbox to\hsize{
\vbox{\noindent CITUSC 01 -- 046   \hfill December 2001 \\
hep-th/0112012 \hfill revised version }}}

\noindent
\vskip1.3cm
\begin{center}

{\Large\bf D-Instantons and Universal Hypermultiplet~\footnote{Supported 
in part by the `Deutsche Forschungsgemeinschaft'}}

\vglue.2in

Sergei V. Ketov~\footnote{On leave from the University of Kaiserslautern, 
Germany} 

{\it Caltech-USC Center for Theoretical Physics\\
     University of Southern California\\
     Los Angeles, CA 90089--2535, USA}
\vglue.1in
{\sl ketov@citusc.usc.edu}

\end{center}
\vglue.2in
\begin{center}
{\Large\bf Abstract}
\end{center}

Quantum non-perturbative geometry of the universal hypermultiplet is 
investigated. We consider the simple case when the D-instantons, originating 
from the Calabi-Yau wrapped D2-branes, preserve a $U(1)\times U(1)$ symmetry 
of the universal hypermultiplet moduli space. The cluster decomposition of 
D-instantons is proved to be valid in a curved spacetime. We find an  
$SL(2,{\bf Z})$ duality-invariant quaternionic solution to the effective NLSM 
metric of the universal hypermultiplet, which is governed by a 
modular-invariant function. This function appears to be {\it the same} 
function found by Green and Gutperle, and describing the modular invariant 
completion of the $R^4$ term by the D-instanton effects in the type-II 
superstring/M-theory. We argue that our solution interpolates between the 
perturbative (large CY volume) region and the superconformal (Landau-Ginzburg)
 region in the universal hypermultiplet moduli space. We also calculate a 
non-perturbative scalar potential in the hyper-K\"ahler limit, when an abelian
 isometry of the universal hypermultiplet moduli space is gauged in the 
presence of D-instantons. 

\newpage

\section{Introduction}

Instanton corrections in compactified M-theory/superstrings are crucial for 
solving the fundamental problems of vacuum degeneracy and supersymmetry 
breaking. Some instanton corrections in the type-IIA superstring theory 
compactified on a {\it Calabi-Yau} (CY) threefold arise due to the Euclidean 
D2-branes wrapped about the CY special (supersymmetric) three-cycles 
\cite{bbs}. Being BPS solutions to the Euclidean ten-dimensional (10d) 
supergravity equations of motion, these wrapped branes are localized in four 
uncompactified spacetime dimensions and thus can be identified with instantons.
 They are called D-instantons. The D-instanton action is essentially given by 
the volume of the supersymmetric 3-cycle on which a D2-brane is wrapped.  The 
supersymmetric cycles (by definition) minimize volume in their homology class. 

At the level of the {\it Low-Energy Effective Action} (LEEA), the effective 
field theory is given by the four-dimendsional (4d), N=2 supergravity with 
some N=2 vector- and hyper-multiplets, whose structure is dictated by the CY 
cohomology, and whose moduli spaces are independent. The hypermultiplet sector
of the LEEA is described by a 4d, N=2 {\it Non-Linear Sigma-Model} (NLSM) 
with a quaternionic metric in the NLSM target (moduli)  space \cite{bw}. Any 
CY compactification gives rise to the so-called {\it Universal Hypermultiplet}
 (UH)  in 4d, which contains a dilaton amongst its field components. 

When the type-IIA supergravity 3-form has a non-vanishing CY-valued 
expectation value, the UH becomes electrically charged. This implies that 
an abelian isometry of the NLSM target (= UH moduli) space is gauged, while 
the UH scalar potential is non-trivial \cite{ps,pot}.

It is of considerable interest to calculate the UH {\it non}-perturbative NLSM
 metric and the associated scalar potential, by including D-instanton 
corrections. The qualitative analysis was initiated by Witten \cite{w1} who 
showed that the D-instanton quantum corrections are given by powers of 
$e^{-1/g_{\rm string}}$, where $g_{\rm string}$ is the type-II superstring 
coupling constant \cite{w1}. The D-instanton induced interactions in the LEEA 
of ten-dimensinal type-II superstrings, and a modular invariant completion of 
the $R^4$-term were found by Green and Gutperle \cite{gg}. These $R^4$-terms 
terms apparently arise from a one-loop calculation in eleven-dimensional 
M-theory \cite{gg2}. The D-brane contributions to the $R^4$-couplings in any
toroidal compactification of type-II superstrings, as well as their 
relation to the Eisenstein series (in eight and seven spacetime dimensions), 
were investigated by Pioline and Kiritsis \cite{pki}. The CY wrapped D-branes 
from the mathematical viewpoint were reviewed by Douglas \cite{doug}.
 
Some explicit D-instanton corrections in the universal sector of the CY 
compactified type-II superstrings were calculated by Ooguri and Vafa \cite{ov},
 though in the hyper-K\"ahler limit when the spacetime gravity is switched 
off. The gravitational corrections are expected to be equally important at 
strong string coupling, while the UH sector is a good place for studying them.
 In particular, Strominger \cite{one} proved the absence of {\it perturbative} 
superstring corrections to the local UH metric provided that the Peccei-Quinn 
type isometries of the classical UH metric, described by the symmetric space 
$SU(2,1)/SU(2)\times U(1)$ \cite{fsh}, are preserved. In our earlier paper 
\cite{plb}  we proposed the procedure for a derivation of the 
non-perturbative UH metric in a curved spacetime. Unfortunately, no explicit 
quaternionic solutions, describing D-instantons, were found in ref.~\cite{plb}.
 In this paper we give such solutions by using some recent advances in 
differential geometry  \cite{cp}. 

We also turn to the gauged version 
of the universal hypermultiplet NLSM, by gauging  one of its abelian 
isometries preserved by D-instantons. This gives rise to the non-perturbative 
scalar potential whose minima determine the `true' vacua in our toy model 
comprising the UH coupled to the single N=2 vector multiplet gauging the UH 
abelian isometry. As is well-known (see, e.g., ref.~\cite{pot}), gauging the 
classical UH geometry  gives rise to the dilaton potential whose minima occur 
outside of the region where the string perturbation theory applies. 
 However, this potential with the run-away behaviour  is not protected against 
instanton corrections, while it is reasonable to gauge only those NLSM 
isometries that are not broken after the D-instanton corrections are included.
 Because of the brane charge quantization, the classical {\it continuous} 
symmetries of the UH metric are generically broken by the wrapped D2-branes 
and the solitionc 5-branes wrapped about the entire CY \cite{bbs}. However,  
when merely D-instantons are taken into account, a continuous abelian symmetry
 of the UH moduli space may survive, while it also makes actual calculations 
possible \cite{nbi}. 

The paper is organized as follows: in sect.~2 we recall a few basic facts 
about the type-II string dilaton and the 4d NLSM it belongs to. In sect.~3 we 
discuss all possible deformations of this NLSM due to D-instantons under the 
condition of unbroken 4d, N=2 local supersymmetry, and give some explicit 
solutions. An $SL(2,{\bf Z})$ modular invariant quaternionic metric solution, 
governed by an order-$3/2$ Eisenstein series, is given in sect.~3 too. 
Sect.~4 is devoted to the gauged version of the UH and its scalar potential in
 the presence of the D-instanton corrected UH metric. Our conclusion is given
in the Abstract. We made all efforts to keep our presentation as simple as 
possible.

\section{Dilaton and NLSM}

In all four-dimensional superstring theories a dilaton scalar $\vf$ is 
accompanied by an axion pseudo-scalar $R$~  
belonging to the same scalar supermultiplet. In the (classical) supergravity 
approximation, their LEEA (or kinetic terms)
 are given by the NLSM whose structure is entirely fixed by duality: 
the NLSM target space is given by the 
two-dimensional non-compact homogeneous 
 space $SL(2,{\bf R})/U(1)$. In the full `superstring theory' 
(including branes) the 
continuous symmetry $SL(2,{\bf R})\cong SO(2,1)
\cong SU(1,1)$ is generically broken to its discrete subgroup $SL(2,{\bf Z})$
 \cite{ht}, 
whereas the local NLSM metric may receive some non-perturbative (instanton) 
corrections \cite{bbs,w1,ov,one,nbi}. 

The $SL(2,{\bf R})/U(1)$-based NLSM can be parametrized in terms of a single 
complex scalar,
$$ A \equiv A_1+iA_2  = R + ie^{-\vf}~,\eqno(2.1)$$
subject to the $SL(2,{\bf R})$ duality transformations
$$ A~\to~A' =\fracmm{aA +b}{cA +d}~,\quad {\rm where}\quad
\left(\begin{array}{cc} a & b \\ c & d \end{array}\right)\in SL(2,{\bf R})~,
\eqno(2.2)$$
with four real parameters $(a,b,c,d)$ obeying the 
 condition $ad-bc=1$. The $SL(2,{\bf R})$ NLSM Lagrangian in the
parametrization (2.1) is given by
$$ \k^2 \Lag(A,\bar{A}) = \fracmm{1}{(A-\bar{A})^2}\pa^{\m}\bar{A}\pa_{\m}A~.
\eqno(2.3)$$
We assume 
 that our scalars are dimensionless. The dimensional 
coupling constant $\k$ of the UH NLSM is proportional to  
the gravitational coupling constant. We assume that $\k^2=1$ 
for notational simplicity.

It is easy to check that the NLSM metric defined by eq.~(2.3) is K\"ahler, 
with a K\"ahler potential
$$ K(A,\bar{A}) = \log(A-\bar{A})~.\eqno(2.4)$$
The $SL(2,{\bf R)}$ transformations (2.2) are generated by constant shifts of 
the axion (T-duality) with
$$ \left(\begin{array}{cc} 1 & 1 \\ 0 & 1 
\end{array}\right)\in SL(2,{\bf R})~,\eqno(2.5)$$
and the S-duality transformation $\left(e^{-\vf}\to e^{+\vf}\right)$ with
$$ \left(\begin{array}{cc} 0 & 1 \\ -1 & 0 \end{array}\right)\in 
SL(2,{\bf R})~.\eqno(2.6)$$

It is worth mentioning that the K\"ahler potential is 
 defined modulo K\"ahler gauge transformations,
$$ K(A,\bar{A}) ~\to~  K(A,\bar{A}) +f(A) +\bar{f}(\bar{A})~,\eqno(2.7)$$
with arbitrary (locally holomorphic) functions $f(A)$. After 
 the field redefinition 
$$ S =i\bar{A}\equiv  e^{-2\f}+i2D~, \eqno(2.8)$$
in terms of a dilaton $\f$ and an axion $D$, 
the  K\"ahler potential (2.4) takes the form
$$ K(S,\bar{S})=  \log(S+\bar{S})~.\eqno(2.9)$$
This parametrization was used, for example, in refs.~\cite{bbs,one,nbi}. 

To connect the K\"ahler potential (2.9) with the standard (Fubuni-Study) 
potential used in the mathematical 
literature, let's make yet another field redefinition,
$$ S=\fracmm{1-z}{1+z}~.\eqno(2.10)$$
The new K\"ahler potential $K(z,\bar{z})$ takes the dual Fubini-Study form
 indeed,
$$ K(z,\bar{z}) = \log(1-\abs{z}^2)~~.\eqno(2.11)$$
The corresponding NLSM Lagrangian is 
$$ -\Lag(\f,D)= (\pa_{\m}\f)^2 +e^{4\f}(\pa_{\m}D)^2~,\eqno(2.12)$$
or   
$$ -4\Lag(\r,t)= \fracmm{1}{\r^2}\left[ (\pa_{\m}\r)^2 
+(\pa_{\m}t)^2\right]~,\eqno(2.13)$$
where we have introduced the new variables
$$ \r=e^{-2\f} \quad{\rm and}\quad t=2D~.\eqno(2.14)$$
The metric of the NLSM (2.13) is conformally flat, it has a negative scalar 
curvature and a manifest isometry, 
due to the $t$-independence of all its components.  

The (complex) one-dimensional K\"ahler potential (2.11) has a natural 
(K\"ahler and dual Fubini-Study type) 
extension to two (complex) dimensions,
$$  K(z_1,z_2,\bar{z}_1,\bar{z}_2)=
\log(1-\abs{z_1}^2 -\abs{z_2}^2)~~,\eqno(2.15)$$
where $(z_1,z_2)\in {\bf C}^2$ are on equal footing inside the ball $B^4$:  
 $\abs{z_1}^2 +\abs{z_2}^2 < 1$.  
The K\"ahler potential (2.15) defines the so-called Bergmann metric 
(in the mathematical literature):
$$ ds^2= \fracmm{dz_1d\bar{z}_1+dz_2d\bar{z}_2}{1-\abs{z_1}{}^2-\abs{z_2}{}^2}
+ \fracmm{(\bar{z}_1dz_1+\bar{z}_2dz_2)(z_1d\bar{z}_1+z_2d\bar{z}_2)}{(1-
\abs{z_1}{}^2-\abs{z_2}{}^2)^2} \eqno(2.16)$$
in the open ball $B^4$. 
Being equipped with the Bergmann metric, the open ball $B^4$ 
 is equivalent to the symmetric 
quaternionic space $SU(2,1)/U(2)$ \cite{besse}. The relation to the UH 
parametrization $(\f,D,C,\bar{C})$ used in 
the physical literature \cite{bbs,fsh,nbi} is given by 
$$  z_1= \fracmm{1-S}{1+S}~,\quad   z_2= \fracmm{2C}{1+S}~~~,\eqno(2.17)$$
where the new complex variable $C$ can be identified with the RR-scalar of 
the UH, whereas another complex
scalar $S$ is now given by  ({\it cf.} eq.~(2.8))
$$ S= e^{-2\f}+i2D +\bar{C}C~.\eqno(2.18)$$

The two complex scalars $(S,C)$ represent all the physical scalars of the 
universal hypermultiplet that also
has a Dirac hyperino as their
 fermionic superpartner. The UH metric defined by eq.~(2.16) is K\"ahler, 
 with a K\"ahler potential 
$$ K(S,\bar{S}, C,\bar{C}) = \log\left(S+\bar{S}-2C\bar{C}\right)~~.
\eqno(2.19)$$
The corresponding 
(bosonic part of)
NLSM Lagrangian of the UH 
 in terms of the scalar fields 
 $(\f,D,C,\bar{C})$ reads 
$$ - \Lag_{\bf FS} = (\pa_{\m}\f)^2 + e^{2\f}\abs{\pa_{\m}C}^2 
+ e^{4\f}(\pa_{\m}D +\fracm{i}{2}\bar{C}
\dvec{\pa_{\m}}C)^2~.~\eqno(2.20)$$
This NLSM metric is diffeomorphism-equivalent to the quaternionic Bergmann 
metric on $SU(2,1)/U(2)$ by our 
construction. At the same time, eq.~(2.20) coincides with the so-called 
{\it Ferrara-Sabharwal} (FS) NLSM 
(in the physical literature) that was derived \cite{fsh} by compactifying the 
10d type-IIA supergravity on a CY 
threefold in the universal (UH) sector down to four spacetime timensions,
 $\m=0,1,2,3$. This means that we can 
identify our field $\f$ with the dilaton used in refs.~\cite{fsh,bbs}.
 We conclude that the FS metric (2.20) is 
completely determined by the duality symmetries of $SU(2,1)/SU(2)\times U(1)$
 \cite{nbi}. The FS metric can be trusted
as long as the string coupling is not strong, 
$g_{\rm string}=e^{\VEV{\f}}=\VEV{1/\sqrt{\r}}$, i.e. for large $\r>0$.  
The variable $\r$ has the physical meaning of the CY space volume
 --- see eq.~(2.14) and ref.~\cite{one}.

\section{D-instantons and quaternionic geometry}  

Quantum non-perturbative corrections generically break all the continuous 
$SU(2,1)$ symmetries 
of the UH classical NLSM down to a discrete subgroup because of 
 charge quantization, even if local N=2 supersymmetry 
in 4d remains unbroken \cite{bbs}. Nevertheless, if we restrict ourselves 
to the special situations when some of the
abelian symmetries of the UH moduli space remain unbroken, 
actual calculations of  
instanton corrections become possible.
 As was demonstrated
by Strominger \cite{one}, there is no non-trivial quaternionic deformation 
of the classical FS metric within the 
superstring perturbation theory when the Peccei-Quinn-type symmetries   
(with three real parameters $(\a,\b,\g)$), 
$$ D\to D+\a~,\quad C\to C+ \g -i\b~,\quad S\to S +2(\g+i\b)C+\g^2+\b^2~,
\eqno(3.1)$$
remain unbroken. However, when {\it some} of these Peccei-Quinn-type 
symmetries (e.g., the one associated with
shifts of the $C$-field) are broken, a calculation of 
the D-instanton contributions is
 possible indeed
 \cite{nbi}. In this
paper we assume that the abelian isometry associated with constant shifts 
of the axionic $D$-field is preserved, as well 
as the $U_C(1)$ duality rotations of the RR-type $C$-field,
$$ D\to D +\a~,\quad C\to e^{i\d}C~,\quad{\rm where}\quad  \d\in [0,2\p]~~.
 \eqno(3.2)$$ 
The isometries (3.2) can hold in the presence of D-instantons \cite{ov,plb}. 
Our considerations in this paper are entirely local, so that in what follows 
the D-instanton corrected UH metric is assumed to be quaternionic (as long as 
local N=2 supersymmetry is preserved) with a single $U(1)$  or 
$U_D(1)\times U_C(1)$ (torus) isometry. The problem now amounts to a 
derivation of non-trivial quaternionic deformations of the Bergmann (or FS) 
metric, which can be physically interpreted as the D-instanton 
contributions, subject to the given abelian isometries.

A generic quaternionic manifold admits three independent {\it almost} complex 
structures $(\tilde{J}_A)\du{a}{b}$, where $A=1,2,3$ 
and $a,b=1,2,3,4$. Unlike the hyper-K\"ahler manifolds, the quaternionic  
 complex structures are {\it not} 
covariantly constant, i.e. they are not integrable to some 
complex structures, due to a non-vanishing NLSM torsion. This torsion is 
induced by 4d, N=2 supergravity because the quaternionic 
condition on the hypermultiplet NLSM target space metric is the direct 
consequence of local N=2 supersymmetry in four spacetime 
dimensions \cite{bw}. As regards real {\it four}-dimensional quaternionic 
manifolds (relevant to the UH), they all have 
{\it Einstein-Weyl\/} geometry of {\it negative\/} scalar curvature 
\cite{bw,besse}, 
$$ W^-_{abcd}=0~,\qquad R_{ab}=-\fracmm{\L}{2}g_{ab}~,\eqno(3.3)$$  
where $W_{abcd}$ is the Weyl tensor, $R_{ab}$ is the Ricci tensor of the 
metric $g_{ab}$, and the constant $\L>0$ is 
proportional to the gravitational coupling constant. The precise value of the
 `cosmological constant' $\L$ in our notation is fixed in eq.~(3.13). 

Since we assume that the UH quaternionic metric has at least one abelian 
isometry, a good starting point is the Tod theorem
\cite{tod} applicable to {\it any} Einstein-Weyl metric of a non-vanishing 
scalar curvature with a Killing vector $\pa_t$. 
According to ref.~\cite{tod}, there exists a parametrization $(t,\r,\m,\n)$ in
 which such metric takes the form
$$ ds^2_{\rm Tod}= \fracmm{1}{\r^2}\left\{ \fracmm{1}{P}(dt+\hat{\Q})^2 
+P\left[ e^u(d\m^2+d\n^2)+d\r^2\right]\right\}~~,\eqno(3.4)$$
in terms of two potentials, $P$ and $u$, and the one-form $\hat{\Q}$ 
in three dimensions $(\r,\m,\n)$. The first 
potential $P(\r,\m,\n)$ is fixed in terms of the second potential $u$ as 
\cite{tod} 
$$ P= \fracmm{3}{2\L} \left(\r\pa_{\r}u-2\right)~~.\eqno(3.5a)$$
The potential $u(\r,\m,\n)$ itself obeys the 3d {\it non-linear} equation 
$$ (\pa^2_{\m}+\pa^2_{\n})u+\pa^2_{\r}(e^u)=0~~,\eqno(3.5b)$$
known as the (integrable) $SU(\infty)$ or 3d Toda equation, whereas 
the one-form $\hat{\Q}$ satisfies the {\it linear} differential 
equation 
$$-d\wedge\hat{\Q} =(\pa_{\n}P) d\m\wedge d\r +(\pa_{\m}P) d\r\wedge d\n 
+\pa_{\r}(Pe^u) d\n\wedge d\m~.\eqno(3.5c)$$   
Some comments are in order. Given an isometry of the quaternionic 
 metric $g_{ab}$ 
with a Killing vector $K^{a}$, 
$$ K^{a;b} + K^{b;a} =0~, \quad K^2\equiv g_{ab}K^aK^b \neq 0~,\eqno(3.6)$$
we can always choose some adapted coordinates, with  all 
 the metric components being 
 independent upon one of the coordinates $(t)$, 
as in eq.~(3.4). We can then 
 plug the Tod Ansatz (3.4) into the Einstein-Weyl 
equations (3.3). It follows \cite{tod} that this 
precisely amounts to the equations (3.5). The proof is straightforward, 
e.g. by the use of Mathematica.

It is worth mentioning that after the conformal rescaling 
$$ g_{ab}\to \r^2 g_{ab}\eqno(3.7)$$
a generic Einstein-Weyl metric of the form (3.4) becomes K\"ahler with the 
vanishing scalar curvature \cite{gau}. After this 
conformal rescaling the Tod Ansatz (3.4) precisely takes the form of the 
standard (LeBrun) Ansatz for scalar-flat K\"ahler 
metrics \cite{lebrun}, 
$$ ds^2_{\rm LeBrun}= \fracmm{1}{P}(dt+\hat{\Q})^2 
+P\left[ e^u(d\m^2+d\n^2)+d\r^2\right]~~,\eqno(3.8)$$
whose potential $u$ still satisfies the 3d Toda equation (3.5b), whereas the 
potential $P$ is a solution to 
$$ (\pa^2_{\m} + \pa^2_{\n})P + \pa^2_{\r}(e^uP)=0~.\eqno(3.9)$$
This equation is nothing but the integrability condition for eq.~(3.5c) 
that holds too.

According to LeBrun \cite{lebrun}, a scalar-flat K\"ahler metric is 
{\it hyper}-K\"ahler if and only if
$$ P \propto \pa_{\r}u~.\eqno(3.10)$$
Given eq.~(3.10), the LeBrun Ansatz reduces to the Boyer-Finley Ansatz 
\cite{bf} for a four-dimensional hyper-K\"ahler metric 
with a rotational isometry \cite{bf}, or to the Gibbons-Hawking Ansatz 
\cite{gh} in the case of a translational (or tri-holomorphic) isometry 
that essentially implies $u=0$ in addition. Both Ans\"atze are well known 
in general relativity (see, e.g., ref.~\cite{book} for
a review). In particular, exact solutions to the Boyer-Finley Ansatz are 
 governed by the same 3d Toda equation, whereas
exact solutions to the Gibbons-Hawking Ansatz \cite{gh}
$$ ds^2_{\rm GH}= \fracmm{1}{P}(dt+\hat{\Q})^2 +P(d\m^2+d\n^2+d\r^2) 
\eqno(3.11)$$
are governed by the {\it linear} equations, $(\pa^2_{\m} + \pa^2_{\n} 
+\pa^2_{\r})P=0$ and $\vec{\de}P+\vec{\de}\times\vec{\Q}=0$,
whose solutions are given by 
 harmonic functions. 
Given another commuting isometry, each of such 
$U(1)\times U(1)$-invariant hyper-K\"ahler metrics is described by 
a harmonic function depending upon two variables, like in ref.~\cite{ov}.

The hyper-K\"ahler geometry arises in the limit when the spacetime N=2 
supergravity decouples, because {\it any} 4d NLSM 
with rigid N=2 supersymmetry has 
 a hyper-K\"ahler metric \cite{fal}. The existence of such approximation 
is dependent upon the validity of eq.~(3.10). Otherwise, the hyper-K\"ahler 
limit may not exist. Given a $U(1)\times U(1)$ isometry
of a hyper-K\"ahler metric, the existence of a translational 
(i.e. tri-holomorphic) isometry does not pose a problem, since there
always exists a linear combination of two commuting abelian isometries 
that is tri-holomorphic \cite{gib}. Some explicit examples of the 
correspondence between four-dimensional hyper-K\"ahler and quaternionic 
metrics were derived in ref.~\cite{iv} from harmonic superspace. 

We are now in a position to rewrite the classical UH metric (2.20) into the 
Tod form (3.4) by using {\it the same} coordinates 
as in eq.~(2.20).  We find
$$ P=1~,\quad e^u=\r~,\quad {\rm and}\quad d\wedge \hat{\Q}
= d\n\wedge d\m~~,\eqno(3.12)$$
which are all agree with eqs.~(3.5a), (3.5b) and (3.5c), respectively. 
Eq.~(3.5a) also implies that
$$ \L=3~.\eqno(3.13)$$

The classical UH metric does not have a hyper-K\"ahler limit because 
$\pa_{\r}u=1/\r$ is not proportional to $P=1$, so that eq.~(3.10)
is not valid. This conclusion is confirmed 
 by direct checking that $\r^2 ds^2_{\rm FS}$ is K\"ahler and scalar-flat,
 but it is 
{\it not} Ricci-flat, and, hence, it is {\it not} hyper-K\"ahler. 
This result does not seem to be very surprising after taking into account 
the 
 fact that the dilaton supermultiplet is dual to the supergravity multiplet 
under the mirror symmetry \cite{one}.   
We can now identify the $\r$ and $t$ coordinates in eq.~(2.14) with the $\r$ 
and $t$ coordinates here, as well as set up
$$ C = \m +i\n~~.\eqno(3.14)$$
The classical UH story is now complete. The non-perturbative UH metrics 
(with instanton corrections) are governed by 
non-separable solutions to the $SU(\infty)$ Toda equation (3.5b) 
with $P\neq 1$ \cite{plb}, and they 
 are very difficult to find \cite{plb,nbi}.

However, we didn't take advantage of the second (linearly independent) abelian
 isometry of the UH metric yet! Given two abelian isometries commuting with 
each other, as in eq.~(3.2), one can write down another Ansatz for the UH 
metric in adapted coordinates where both isometries are manifest (i.e. in 
terms of some potentials depending upon {\it two} coordinates only), and 
then impose the Einstein-Weyl conditions (3.3). Surprisingly enough, 
this programm was successfully accomplished in the mathematical
literature only recently by Calderbank and Petersen \cite{cp}.   

The main result of ref.~\cite{cp} is the theorem that {\it any} 
four-dimensional quaternionic metric (of a non-vanishing scalar 
curvature) with two linearly independent Killing vectors can be 
written down in the from 
$$\eqalign{ 
ds^2_{\rm CP} ~=~ &  \fracmm{4\r^2(F^2_{\r}+F^2_{\h})-F^2}{4F^2}\,
\left(\fracmm{d\r^2+d\h^2}{\r^2}\right) \cr 
 & + \fracmm{ [(F-2\r F_{\r})\hat{\a}-2\r F_{\h}\hat{\b} ]^2 +[-2\r F_{\h}
\hat{\a}
+(F+2\r F_{\r})\hat{\b}]^2 }{F^2[4\r^2(F^2_{\r}+F^2_{\h})-F^2] }~,\cr}
\eqno(3.15)$$
in some local coordinates $(\r,\h,\q,\j)$ inside an open region of the 
half-space $\r>0$, where $\pa_{\q}$ and $\pa_{\j}$ are the two Killing 
vectors, the one-forms $\hat{\a}$ and $\hat{\b}$ are given by
$$ \hat{\a}= \sqrt{\r}\,d\q\quad {\rm and}\quad \hat{\b}=\fracmm{d\j 
+\h d\q}{\sqrt{\r}}~~,\eqno(3.16)$$
and, most importantly, the whole metric (3.15) is governed by the function 
$F(\r,\h)$ that is a solution to the {\it linear} differential equation
$$ \left(\pa^2_{\r}+\pa^2_{\h}\right)F = \fracmm{3}{4\r^2}F~~.\eqno(3.17)$$

Some comments are in order. 

First, it is fairly straightforward (e.g., by using Mathematica) to verify 
that the {\it Calderbank-Petersen} (CP) Ansatz (3.15) does satisfy the 
fundamental Einstein-Weyl equations (3.3) under the conditions (3.16) and 
(3.17), so that the metric (3.15) is quaternionic indeed. Moreover, this 
calculation is also useful to verify that the metric (3.17) is of 
{\it negative} scalar curvature provided that \cite{cp} 
$$ 4\r^2(F^2_{\r}+F^2_{\h}) > F^2>0~~.\eqno(3.18)$$

Second, the field redefinition \cite{iv}
$$ G= F\sqrt{\r} \eqno(3.19)$$
allows one to rewrite the CP Ansatz (3.15) to the form
$$-ds^2= G^{-2}\left\{ \fracmm{1}{P}(d\j +\hat{\Q})^2 + Pd\g^2\right\}~,
\eqno(3.20)$$
where \cite{iv}
$$ P= 1- \fracmm{GG_{\r}}{\r(G^2_{\r}+G^2_{\h})}~,\quad \hat{\Q}=
\left(\fracmm{GG_{\h}}{G^2_{\r}+G^2_{\h}}-\h\right)d\q~,\eqno(3.21)$$
and
$$ d\g^2 \equiv \r^2 d\q^2 + (G^2_{\r}+G^2_{\h})(d\r^2+d\h^2)~~.\eqno(3.22)$$
The Ansatz (3.20) is similar to the Tod Ansatz (3.4), while it allows us to 
identify the $G$ function (3.19) with the Tod coordinate $\r$ in eq.~(3.4). 
Plugging eq.~(3.19) into eq.~(3.17) yields the linear differential equation on
 $G(\r,\h)$ \cite{iv}:
$$ \left(\pa^2_{\r}+\pa^2_{\h}\right)G = \fracmm{1}{\r}\pa_{\r}G~~.
\eqno(3.23)$$
Unfortunately, eq.~(3.22) does not seem to imply a direct relation between the
 Toda potential $u$ in eq.~(3.4) and a function $F$ in eq.~(3.17) since yet 
another reparametrization is needed to put the Ansatz (3.20) into the Tod form
 (3.4).

Third, and, perhaps, most importantly, the linear equation (3.17) means that 
$F$ is a local eigenfunction, with eigenvalue $3/4$, of the two-dimensional 
Laplace-Beltrami operator on the hyperbolic plane $\ch^2$ with the metric 
$$ ds^2_{\ch}= \fracmm{1}{\r^2}( d\r^2 +d\h^2)~. \eqno(3.24)$$  
Unlike the non-linear Toda equation (3.5b), the linearity of eq.~(3.17) allows
 a superposition of any two solutions to form yet another solution. In 
physical terms, this amounts to the cluster decomposition of D-instantons. 
The validity of such decomposition is not obvious in a curved spacetime.

Though we cannot identify a dilaton in the full moduli space of the UH 
(the NLSM of the UH has general coordinate invariance in its target space), we
 can do it in the perturbative region where the string coupling is weak, i.e.
at {\it large} $\r\to +\infty$, which implies
$$ G\propto \r^2\to +\infty~,\quad P\to const. \quad {\rm and}\quad F\to 
 \r^{3/2}~~,\eqno(3.25)$$   
where we have used eqs.~(3.17), (3.19) and (3.21). 

A simple solution to eqs.~(3.17) and (3.23) outside of the perturbative region
 is given by
$$G=1 \quad {\rm and} \quad F_1=\fracmm{1}{\sqrt{\r}}~~~. \eqno(3.26)$$
This solution looks like an `instanton' solution but it implies 
$4\r^2(F^2_{\r}+F^2_{\h})-F^2=0$  that is incompatible with eq.~(3.18). 
The `multi-instanton' solutions do exist \cite{cp}. However, first, we have 
to impose some more physical requirements on them.

First, we expect the exact (non-perturbative) UH moduli space metric to be 
{\it complete} inside some four-dimensional ball $B^4$ \cite{pro} ---  
by analogy with the exact Seiberg-Witten-type solutions in the 
non-perturbative N=2 supersymmetric gauge field theories (see, e.g., 
ref.~\cite{book} for a review).

Second, the full UH metric should also respect the known topological 
`boundary conditions': in the perturbative region it should reduce to the 
standard (Bergmann or Ferrara-Sabharwal) classical metric up to 
diffeomorphisms, while it should also possess the UV fixed point 
(or a conformal infinity \cite{leb2}) at some point of $B^4$ outside of the 
perturbative domain where one expects the N=2 superconformal field theory 
(or Landau-Ginzburg) description to be valid \cite{doug}. 

Third, because the D-instantons (i.e. the D2-branes wrapped about the special 
3-cycles of CY) are supposed to be the origin of non-perturbative corrections 
to the UH metric, we should expect the dependence of this metric upon the 
RR-type $\h$-variable to be periodic. Indeed, given an Euclidean D2-brane 
wrapped $m$-times around $S^3$ in CY, it couples to the RR expectation value 
on $S^3$ and thus produces a factor of $\exp(2\p im\h)$ --- {\it cf.} 
ref.~\cite{ov}. We should, therefore, search for a solution to eq.~(3.17) in 
the form of a D-instanton sum that is periodic in $\h$ \cite{plb}. 

Fourth, a discrete $SL(2,{\bf Z})$ duality symmetry is supposed to be the 
exact symmetry of the full type-II `supertring' theory  (including branes). 
Hence, it must be a symmetry of our UH effective metric solution. 

To the best of our knowledge, no solutions with all the above-mentioned 
features are known. Nevertheless, it is known how to construct exact 
`multi-centre' solutions to eq.~(3.17) with a {\it finite} instanton number 
$m>1$ \cite{gl,bgnp}. These solutions were originally found by using the 
quaternionic-K\"ahler quotients of the $4(m-1)$-dimensional quaternionic 
projective space $HP^{m-1}$ by an $(m-2)$-torus (i.e. by an 
$(m-2)$-dimensional family of commuting Killing vectors) with $m>1$ 
\cite{bf,bgnp}. In the case relevant for our investigation, the quaternionic 
projective plane $HP^{m-1}$ should be replaced by the non-compact quaternionic
 hyperboloid, $H^{m-1}\to H^{p-1,q}$, with $(p,q)=(m-1,1)$.

In terms of the CP description \cite{cp} of the four-dimensional quaternionic 
metrics with torus isometry, governed by eq.~(3.17), the `multi-instanton' 
metric solutions are described by the following simple solution to the linear 
equation (3.17) on $\ch^2$: 
$$ F_m(\r,\h) = \sum_{k=1}^{m} \fracmm{\sqrt{a^2_k\r^2 
+(a_k\h -b_k)^2}}{\sqrt{\r}} \eqno(3.27)$$
with some real moduli $(a_k,b_k)$. Since the superposition principle applies, 
it is easy to check that eq.~(3.27) is a solution to eq.~(3.17)  indeed. Each 
additive contribution to the right-hand-side of eq.~(3.27) is just a simple 
generalization of the `basic' solution (3.26) corresponding to  $a=0$ and 
$b=1$.  When the hyperbolic plane $\ch^2$ is mapped onto an open disc $D$, the
 `positions' of instantons are given by the marked points on the boundary of 
this disc where the torus action has its fixed points. The `twistors' (in the 
terminology of ref.~\cite{cp}) $\{a_k,b_k\}^{m}_{k=1}$ form the 
$2m$-dimensional vector space where the three-dimensional $SL(2,{\rm R})$ 
duality group naturaly acts. In addition, the solutions $F$ are merely defined 
modulo an overall real factor. Hence, the total (real) dimension of the 
D-instanton moduli space $\cm_{m}$ (of a finite instanton number $m$) is given
 by \cite{cp} $$ {\rm dim}\, \cm_m = 2m-{\rm dim}\,SL(2,{\bf R})-1 = 2m-4~.
\eqno(3.28)$$ 
As is clear now, the whole moduli space (for all $m$) is infinite dimensional,
 in agreement with the LeBrun theorem \cite{leb3}.

Many explicit examples of the four-dimensional quaternionic metrics in the 
case of $m=2$ and $m=3$ can be found, e.g., in ref.~\cite{cp}. These examples 
include, in particular, the quaternionic-K\"ahler extensions of generic 
four-dimensional hyper-K\"ahler metrics with two centers and $U(1)\times U(1)$ 
isometry (like Taub-NUT and Eguchi-Hanson) \cite{iv}. In the case of $m=2$ one
 finds only non-interesting (hyperbolic) metrics that have nothing to do with 
instantons or monopoles. The most general solution in the case of $m=3$ reads
 \cite{cp}
$$ F_3(\r,\h)=\fracmm{1}{\sqrt{\r}}+\fracmm{(b+c/q)\sqrt{\r^2
+(\h+q)^2}}{\sqrt{\r}}+
\fracmm{(b-c/q)\sqrt{\r^2+(\h-q)^2}}{\sqrt{\r}}~~,\eqno(3.29)$$
where $(b,c)$ are two real (non-negative) moduli and $q^2=\pm 1$. Amongst the 
solutions (3.29) there are the ones that do possess the physically important 
features, such as \cite{ped,gl,gal,ag,cp}
\begin{itemize}
\item completeness in certain domains,
\item negative scalar curvature,
\item existence of the UV fixed point (or a conformal infinity), 
\item smoothness, i.e. no unremovable singularities, 
\item equivalence to the classical metric at some special values of the moduli.
\end{itemize} 
The holographic principle may also apply on the conformal boundary 
\cite{ket4}. We have, therefore, good reasons to expect all these  features 
to be valid for higher $m$ too. 

The periodicity condition with respect to the RR-type coordinate $\h$,
$$ F(\r,\h) = F(\r,\h+1)~, \eqno(3.30)$$
is non-trivial since it cannot be true for any finite value of $m$. However, 
eq.~(3.30) may be satisfied by the {\it infinite} D-instanton sum, {\it viz.}
$$ F(\r,\h) = \fracmm{1}{\sqrt{\r}} + \sum^{+\infty}_{n=-\infty} \abs{a_n}  \,
\fracmm{\sqrt{\r^2+(\h+n)^2}}{\sqrt{\r}}=
\fracmm{1}{\sqrt{\r}} + \sum^{+\infty}_{n=-\infty} \abs{a_n}  \,
\sqrt{\r +\fracmm{(\h+n)^2}{\r}}\eqno(3.31)$$
whose moduli $\{a_n\}$ are supposed to guarantee convergence of the infinite 
series. We conclude that {\it all}\/ D-instanton (winding) numbers have to be 
present in the non-perturbative corrections to the UH metric. 

In fact, the full UH solution to eq.~(3.17), which describes all D-instanton
corrections, should also respect the non-perturbative $SL(2,{\bf Z})$ duality
(sect.~2),
$$ \t ~\to~ \fracmm{a\t +b}{c\t +d}~,\quad (a,b,c,d) \in {\bf Z}~,\quad
ad-bc=1~,\eqno(3.32)$$
where we have introduced the complex coordinate $\t$,
$$ \t = \t_1 + i\t_2\equiv \h + i\r~~.\eqno(3.33)$$
It is, therefore, natural to search for a non-holomorphic solution 
$F(\t,\bar{\t})$ amongst the modular functions $f_s(\t,\bar{\t})$   of order
$s$, which are defined by the Eisenstein series \cite{ter}
$$ f_s(\t,\bar{\t}) = \sum_{(p,n)\neq (0,0)}\left(\fracmm{\t_2}{\abs{p
+n\t}^2}\right)^s=\sum_{(p,n)\neq (0,0)}\fracmm{\r^s}{[p^2 +n^2(\h^2+\r^2)
+2np\h]^s}~~,\eqno(3.34)$$
and obey the eigenvalue equation
$$ \t^2_2(\pa^2_{\t_1}+\pa^2_{\t_2})f_s(\t,\bar{\t})=s(s-1)f_s(\t,\bar{\t})~.
\eqno(3.35)$$

Equation (3.17) exactly coincides with eq.~(3.35) provided that $s(s-1)=3/4$ or
$$ s=3/2~~.\eqno(3.36)$$
Thus we conclude that the UH solution to eq.~(3.17) is proportional to 
$ f_{3/2}(\t,\bar{\t})$,
$$ F(\t,\bar{\t})\propto  f_{3/2}(\t,\bar{\t})=\sum_{(p,n)\neq (0,0)}
\fracmm{\t_2^{3/2}}{\abs{p+n\t}^3}~~.\eqno(3.37)$$

The modular-invariant function $f_{3/2}(\t,\bar{\t})$ is precisely {\it 
the same} function that describes multi-instanton contributions to the 
$R^4$ terms in the ten-dimensional type-II superstrings \cite{gg}, which are
due to instantons of discrete energy $p$ and discrete charge $n$. The
solution (3.37) is simply related to the $K_1$ Bessel function \cite{gg},
$$F(\r,\h)\propto f_{3/2}(\t,\bar{\t})= 2\z(3) \r^{3/2} +\fracmm{2\p^2}{3}
\r^{-1/2} + 8\p\r^{1/2}\sum_{m\neq 0 \atop n\geq 1} \abs{\fracmm{m}{n}}
e^{2\p imn\h}K_1(2\p\abs{mn}\r)~,\eqno(3.38)$$
where $\z(3)= \sum_{m>0}(1/m)^3$. The asymptotical expansion of the function
(3.38) in the perturbative (large $\r$) region is given by  \cite{gg}
$$\eqalign{
 f_{3/2}(\t,\bar{\t}) = &  2\z(3)\r^{3/2} +\fracmm{2\p^2}{3}\r^{-1/2} 
+4\p^{3/2}\sum_{m,n\geq 1}\left(\fracmm{m}{n^3}\right)^{1/2}\left[
e^{2\p i mn(\h+i\r)}+e^{-2\p i mn(\h-i\r)}\right]\times \cr
  & \times \left[ 1 + \sum^{\infty}_{k=1}\fracmm{\G(k-1/2)}{\G(-k-1/2)}\,
\fracmm{1}{(4\p mn\r)^k}\right]~~,\cr}\eqno(3.39)$$
while it is just the sum of tree level, one-loop, and instanton contributions
indeed. The one-loop correction has no local meaning since it can be removed 
by a NLSM field redefinition \cite{one}. Since the function (3.39) is periodic
 in $\h$, it should be possible to rewrite it into the form (3.31).

The exact UH quaternionic metric governed by the solution (3.37) via eq.~(3.15)
 does not depend upon details of the CY moduli space, as it may have been 
expected in the {\it universal} sector of CY compactification. The discovered 
relation to the earlier results \cite{bbs,gg,gg2,pki,ov} about D-instantons is 
important for justifying the consistency of our approach.

It is also worth mentioning that to be quaternionic does {\it not} 
automatically mean to be K\"ahler. Though the classical UH metric
is quaternionic-K\"ahler (sect.~2), this does not apply to the 
instanton-corrected quaternionic metrics discussed in this section. 
This observation also implies that our results for the UH coupled to N=2 
supergravity cannot be rewritten to the N=1 supergravity form without 
truncation, since N=1 local supersymmetry in 4d requires the NLSM metric 
to be K\"ahler.

\section{The instanton-induced scalar potential}

When the UH is electrically charged, its scalar potential becomes non-trivial 
(we do not consider any magnetic charges here). This happens because of 
{\it gauging} of an abelian isometry in the UH moduli space. Abelian gauging 
in the {\it classical} UH target space was discussed in great detail
in ref.~\cite{pot} -- see also refs.~\cite{mich,tv,dall}. Since the 
D-instantons are supposed to preserve the $U(1)\times U(1)$ isometry of the 
classical UH moduli space, it is quite natural to gauge a $U(1)$ part of it
 in the presence of the D-instanton quantum corrections, in order to generate 
a {\it non-perturbative} UH scalar potential. The fixed points (zeroes) of the
 UH scalar potential determine new vacua in type-II string theory. Gauging an 
abelian isometry introduces an extra N=2 vector gauge multiplet into our model
 of the UH. In 4d, N=2 supergravity it is the quaternionic NLSM metric and its 
Killing vector that fully determine the corresponding scalar potential. 
Unfortunately, the literature about the hypermultiplet scalar potentials in 
N=2 supergravity is rather confusing (or very complicated, at least), so we 
begin with the case of 4d, {\it rigid} N=2 supersymmetry that is well 
understood \cite{town,sakai}. 

Any hyper-K\"ahler N=2 NLSM in 4d can be obtained from its counterpart in 6d 
by dimensional reduction. No scalar potential for hypermultiplets
is possible in 6d. Hence, a non-trivial scalar potential can only be generated
 via a Scherk-Schwarz-type mechanism of dimensional reduction 
with a non-trivial dependence upon extra spacetime coordinates, like 
\cite{town,sakai}
$$ \pa_4 \f^a = K^a(\f)~,\quad {\rm and} \quad \pa_5 \f^a =0~,\eqno(4.1)$$
where $K^a(\f)$ is a Killing vector in the NLSM target space with a 
hyper-K\"ahler metric $g_{ab}(\f)$ parametrized by four real scalars 
$\f^a$ and $a=1,2,3,4$. The Scherk-Schwarz procedure is consistent with rigid 
N=2 supersymmetry if and only if the Killing vector $K^a(\f)$ represents a 
translational (or tri-holomorphic) isometry, while there is always one such 
isometry in the case of an $U(1)\times U(1)$ symmetric hyper-K\"ahler metric. 
 Upon the dimensional reduction (4.1) down to 4d, the 6d NLSM kinetic terms 
produce the scalar potential 
$$ V(\f) = \fracmm{1}{2}\,g_{ab}K^aK^b \equiv \fracmm{1}{2}K^2 \eqno(4.2)$$
that is just given by half of the Killing vector squared.

Given a four-dimensional hyper-K\"ahler metric with a triholomorphic isometry,
 we are in a position to use the Gibbons-Hawking Ansatz (3.11) 
where this isometry is manifest, with $K^a=(1,0,0,0)$ and $\f^a=(t,\m,\n,\r)$.
  Equation (4.2) now implies  
$$ V = \fracmm{g_{tt}}{2}=\fracmm{1}{2}P^{-1}~~,\eqno(4.3)$$
where $P(\vec{X})$ is a harmonic functon of $\vec{X}=(\m,\n,\r)$.  For example,
 the (Gibbons-Hawking) multi-centre hyper-K\"ahler metrics 
are described by the harmonic function \cite{gh}
$$  P(\vec{X}) =\sum^m_{k=1} \fracmm {1}{\abs{\vec{X}-\vec{X}_k}}~~,
\eqno(4.4)$$
where the moduli $\vec{X}_i$ denote locations of the centers. The 
corresponding scalar potential (4.3) is
non-negative, while its absolute minima occur precisely at the fixed points 
where the harmonic function (4.4) diverges. Since $V=0$ at
these points, N=2 supersymmetry remains unbroken in all of these vacua. 
It is worth mentioning that the vacua are independent upon the NLSM 
parametrization used (the fixed points are mapped into themselves  under 
the NLSM reparametrizations). 

In the case of the UH with 4d, {\it local} N=2 supersymmetry (i.e. coupled to 
N=2 supergravity), we have to deal with a quaternionic NLSM metric having 
a gauged abelian isometry. First, the Gibbons-Hawking Ansatz (3.11) is to be 
replaced by the Tod Ansatz (3.4) \cite{plb}. Second, we have also take into 
account the presence of the abelian gauge N=2 vector multiplet whose complex 
scalar component $(a)$  is going to enter the NLSM scalar potential too. 
As was demonstrated in refs.~\cite{tv,dall}, the scalar potential in the 
gauged N=2 supergravity appears to be a very natural generalization of the 
scalar potential (4.2) in the absence of N=2 supergravity, 
$$ V = \fracmm{1}{{\rm Im}[\t(a)]}\,\fracmm{1}{2}K^2 =  
\fracmm{1}{{\rm Im}[\t(a)]}\,\fracmm{P^{-1}}{2\r^2}~~,\eqno(4.5)$$
where $\t(a)$ is the function governing the kinetic terms of the N=2 vector 
multiplet. We have used eq.~(3.4) in the second equation (4.5).

The standard way of deriving the scalar potential in the gauged N=2 
supergravity uses the local N=2 supersymmetry transformation laws of the
fermionic fields (gauginos, hyperinos and gravitini) \cite{cgp,mich}. 
The contribitions of gauginos and hyperinos are always positive, whereas
the contribution of gravitini is negative. The recent results of 
ref.~\cite{tv,dall} imply that the negative (gravitini) contribution to the
scalar potential cancels against the positive contributions due to the matter 
fermions (gaugino and hyperino) in the gauged N=2 supergravity.
This is not the case in the gauged N=1 supergravity theories \cite{n1}. 
Hence, it may not be possible to rewrite an N=2 gauged supergravity
theory in the N=1 locally supersymmetric form without truncations ({\it cf.} 
our remarks at the end of sect.~3).

Because of unitarity of the N=2 supergravity theory, effectively describing 
the unitary CY-compactifed  theory of type-II superstrings, there should be no 
ghosts in the N=2 vector multipet sector too, so that we should have  
$${\rm Im}[\t(a)]>0~~.\eqno(4.6)$$   
Being interested in the vacua of the effective N=2 supergravity theory, which 
are determined by the minima of its scalar potential (4.5), 
we do not need to know the function $\t(a)$ explicitly -- eq.~(4.6) is enough. 

In the classical approximation for the UH metric, eq.~(3.12) tells us that 
$P=const.$ This immediately gives rise to the run-away behaviour of the 
potential (4.5) with its absolute minimum at $\r=\infty$, in agreement with 
refs.~\cite{pot,dall}. This run-away solution is, of course, physically 
unacceptable because it implies the infinite CY volume i.e. a 
decompactification, as well as the `infinite' string coupling. One may hope 
that the use of the full (non-perturbative) UH metric may improve the scalar 
potential behaviour, because the D-instanton corrections imply that 
$P\neq const.$ (see sect.~3).  Unfortunately, finding an exact potential in 
this case amounts to solving the (non-linear, partial differential) Toda 
equation (3.5b), since the $P$-function is governed by the Toda potential via 
eq.~(3.5a). Though the 3d Toda equation is known to be integrable, it is
notorously difficult to find its explicit (non-separable) solutions.

When N=2 supergravity is switched off (after gauging), we can take a solution 
for the $P$-function in the hyper-K\"ahler limit \cite{ov},
$$\eqalign{
 P(\r,\h)& ~=~ \fracmm{1}{4\p}\sum_{n=-\infty}^{+\infty} \left( 
\fracmm{1}{ \sqrt{\r^2/g_{\rm string}^2 +(\h+n)^2}}-
\fracmm{1}{\abs{n}}\right)+const. \cr
    & ~=~ \fracmm{1}{4\p}\log\left( \fracmm{1}{\r^2}\right)
+\sum_{m\neq 0}\fracmm{1}{2\p}e^{2\p im\h}K_0\left(2\p
\fracmm{\abs{m\r}}{g_{\rm string}}\right)~,}
\eqno(4.7)$$ 
where $K_0$ is the modified Bessel function, and we have re-introduced the 
dependence upon the string coupling constant $g_{\rm string}$ for 
reader's convenience.  The conjectured $U(1)\times U(1)$ symmetry  of the UH 
metric in the form (3.11) and the Poisson resummation formula were used in 
deriving eq.~(4.7) --- see ref.~\cite{ov}. In the perturbative region 
(large $\r$) the asymptotic expansion of the Bessel function in eq.~(4.7) 
yields the infinite D-instanton sum \cite{ov}
$$\eqalign{
 P(\r,\h)~=~&\fracmm{1}{4\p}\log \left( \fracmm{1}{\r^2}\right) +
\sum_{m=1}^{\infty} \exp \left(-\,\fracmm{2\p\abs{m\r}}{g_{\rm string}}\right)
\cos(2\p m\h)\cr
~& \times \sum_{n=0}^{\infty}\fracmm{\G(n+\fracm{1}{2})}{\sqrt{\p}n!
\G(-n+\fracm{1}{2})}\left(\fracmm{g_{\rm string}}{4\p\abs{m\r}}
\right)^{n+\frac{1}{2}} ~~~~.\cr}\eqno(4.8)$$
The $\exp{(-1/g_{\rm string})}$ type dependence of the solution (4.8) 
agrees with the general expectations \cite{w1} so that eq.~(4.8) describes the
 D-instantons indeed. 

We conclude that the non-perturbative vacua of our toy model for the 
electrically charged UH in the presence of D-instantons are given by poles of 
the $P$-function defined by eqs.~(3.4) and (3.5). In the hyper-K\"ahler limit,
 the vacua are given by the fixed points of the D-instanon function (4.7).

\section*{Acknowledgements}

The author would like to thank the Caltech-USC Center of Theoretical Physics 
for kind hospitality extended to him during a preparation of this paper. 
Useful discussions with K. Behrndt, S. J. Gates, Jr., E. Kiritsis, W. Lerche, 
N. Sakai and B. de Wit are gratefully acknowledged.

\end{document}